\documentclass[preprint,authoryear,12pt]{elsarticle}

\usepackage{amssymb}

\journal{Planetary and Space Science}

\begin{document}

\begin{frontmatter}



\title{A framework for resolving the origin, nature and evolution of the diffuse interstellar band carriers?}


\author{Anthony P. Jones}

\address{Institut d'Astrophysique Spatiale (IAS), CNRS and Universit\'e Paris Sud, UMR8617, 91405 Orsay, France. e-mail: {\tt Anthony.Jones@ias.u-psud.fr}}

\begin{abstract}
The carriers of the diffuse interstellar bands (DIBs) still remain an unknown commodity. 
Both dust and molecules have been suggested as carriers but none proposed have yet been able to explain the nature and the diversity of the DIBs. 
Hence, it is perhaps time to review the problem in terms of the intermediate-sized nano-particles. 
It is here proposed that the DIB carriers are the nm-sized and sub-nm-sized products of the UV photo-fragmentation of hydrogenated amorphous carbon grains, a-C(:H), and their heteroatom-doped variants, a-C:H:{\it X} (where $X$ may be O, N, Mg, Si, Fe, S, Ni, P, \ldots).
An interstellar hydrogenated amorphous carbon dust evolutionary framework is described within which a solution to the age-old DIB problem could perhaps be found. 
\end{abstract}


\end{frontmatter}



\section{Introduction}
\label{sect_intro}

The diffuse interstellar band (DIB) problem has now been with us for almost 100 years and we still appear to be no nearer discovering the carriers of the DIBs. Hence, it is perhaps opportune to begin to think outside of the box for a solution to this long-standing conundrum. 
During the course of DIB history our ideas have oscillated back and forth between molecular and dust carriers but perhaps the solution lies in the intermediate domain between these two cases, {\it i.e.}, in the physics of the nano-particles that lie at the molecule-particle interface?

\section{The nature of the DIB carriers}
\label{sect_DIBs}

\cite{1995ARA&A..33...19H} gives a comprehensive review of all of the key DIB and DIB-related issues. 
For a more recent and the state-of-the-art view of this subject the reader is referred to the, shortly to be published, proceedings of the latest meeting dedicated to DIBs (IAU 297, ``The Diffuse Interstellar Bands'', Noordwijkerhout, 20-24 May 2013). 
The following provides an incomplete list of the currently-known observational constraints on the DIBs observed in the interstellar medium (ISM) and their carriers. The DIBs:  

\begin{itemize} 
\item are generally associated with the diffuse rather than the dense ISM,  
\item correlate with the dust extinction, $E(B-V )$,
\item show a weak positive correlations with the 217\,nm UV bump, 
\item show a weak negative correlation with the FUV extinction, 
\item are broader than atomic or molecular lines,   
\item are weaker in strong UV radiation fields,  
\item that are broader are less sensitive to UV radiation, 
\item correlate with small molecules/radicals (but less well than with dust), 
\item are environment-dependent and   
\item generally do not correlate well with each other.
\end{itemize}
This list is certainly incomplete and the logical connections between the various constraints on the nature of the DIB carriers have yet to be melded into a coherent model. 

\section{Amorphous hydrocarbon dust evolution in the ISM}
\label{sect_evolution}

In the condensed matter community it has long been known that hydrogenated amorphous carbons, a-C:H, darken with ultraviolet (UV) irradiation and thermal annealing to form hydrogen-poorer amorphous carbons, a-C \citep[{\it e.g.}, see][and references therein]{2012aA&A...540A...1J,2012bA&A...540A...2J,2012cA&A...542A..98J}. This processing leads to a loss of hydrogen atoms from the structure and an evolution from aliphatic-rich towards aromatic-rich materials, {\it i.e.}, a-C:H $\rightarrow$ a-C, associated with a closing of the band gap, $E_{\rm g}$, of the material. 
This evolution of the band gap is indeed the key to explaining the observed variations in the a-C(:H) optical and structural properties. 

The interesting physical and optical properties of the full suite of (hydrogenated) amorphous carbon materials \{ a-C:H : a-C \}, or a-C(:H) for short, have resulted in quite some interest in them as a model for the solid carbonaceous matter in the ISM \citep[{\it e.g.},][]{1990QJRAS..31..567J,1995ApJ...445..240D,1997ApJ...482..866D,2004A&A...423..549D,2004A&A...423L..33D,2005A&A...432..895D,2008A&A...490..665P,2008A&A...492..127S,2009ASPC..414..473J,2010A&A...519A..39G,2011A&A...529A.146G,2011A&A...525A.103C,2013A&A...555A..39J,2013A&A...xxx...xxx}, in circumstellar media \citep[{\it e.g.},][]{2003ApJ...589..419G,2007ApJ...664.1144S} and in the Solar System \citep[{\it e.g.},][]{2011A&A...533A..98D}.
In the ISM a-C(:H) grains can undergo rather complex, size-dependent evolution arising, principally, from ultraviolet (EUV-UV) photon absorption leading to photo- and/or thermal-processing \cite[{\it e.g.},][]{2009ASPC..414..473J,2012aA&A...540A...1J,2012bA&A...540A...2J,2012cA&A...542A..98J}  and incident ion and electron collisions in shock waves and in a hot gas \cite[{\it e.g.},][]{2010aA&A...510A..36M,2010bA&A...510A..37M,2012A&A...545A.124B}. 

In the ISM the principal process that drives the transformation of the structure and composition of a-C(:H) dust is UV-EUV photo-processing and the relevant time-scales for a-C(:H) photo-processing will be composition- and size-dependent and are of the order of a $10^5-10^6$ years for a-C(:H) nano-particles \citep[{\it e.g.},][]{2012cA&A...542A..98J}. 
Within this framework for the physical, structural and optical properties of a-C(:H) materials \citep{2012aA&A...540A...1J,2012bA&A...540A...2J,2012cA&A...542A..98J} it has been possible to construct an interstellar dust model that, for the first time, coherently takes into account the evolution of these materials within the astrophysical context \citep{2013A&A...xxx...xxx,2013A&A...555A..39J}.  

Principal among the photo-processes that drive a-C(:H) evolution in the ISM in general, and in photo-dissociation regions (PDRs) in particular, is their photo-fragmentation, which derives from the UV-photo-dissociative rupture of the structure, a process that is enhanced with respect to bulk materials by the small (nm) particle sizes. 
There is strong observational evidence for the destruction of the IR emission band carriers in intense radiation fields \citep[{\it e.g.}, $G_0 > 10^3$,][where $G_0$ is the interstellar radiation field intensity normalised to that of the Solar neighbourhood]{1994A&A...284..956B}. More recently, \cite{2012A&A...542A..69P} found evidence for small carbonaceous grain photo-fragmentation in PDRs.  
Their study shows that the photo-fragmentation of small carbon grains into IR emission band carriers occurs in relatively dense and UV irradiated PDRs (H atom density $n_{\rm H} = 10^2-10^5$ cm$^{-3}$, $G_0 = 10^2-5 \times 10^4$, {\it i.e.}, $0.5 < G_0/n_{\rm H} < 1.0$). 
It is therefore clear from the observational evidence that carbonaceous nano-particle photo-fragmentation and destruction does occur in response to EUV-UV photon irradiation in intense radiation fields. 

It is also evident from observations that (photo-)fragmentation can play a role in the more diffuse ISM. For instance, 
in their carbon depletion study \cite{2012ApJ...760...36P} find that the FUV extinction gradually decreases with decreasing density, when normalised by $E(B-V)$, which they interpret as an indication of the preferential fragmentation of small grains in the diffuse ISM. However, this effect can also be explained by the progressive a-C:H to a-C transformation in the diffuse ISM, where the FUV dust cross-section decreases and the UV bump strength increases as the a-C(:H) band gap decreases \citep[see Fig.~6 in][]{2013A&A...xxx...xxx}.
The work of \cite{2012ApJ...760...36P} therefore suggests that carbonaceous dust undergoes significant processing in the neutral ISM and lends support to the idea that dust mass variations in the small carbonaceous grain population can be driven by photo-processing and/or photo-fragmentation in low density regions. 

Thus, photo-fragmentation of a-C(:H) can lead to the liberation of hydrocarbon fragments, such as arophatics \citep{2012ApJ...761...35M}, which consist of aromatic clusters linked by aliphatic and olefinic bridging structures. 
In this context, a-C(:H) nano-particles are particularly susceptible to disruption by EUV-UV photon driven processes that drive arophatic cluster formation and lead to the eventual demise of the fragmentation products as they break down into and liberate daughter molecule/radical/ion species such as H$_2$, C$_2$, CN, CCH, C$_3$, C$_3$H$^+$, c-C$_3$H$_2$ and C$_4$H \citep[{\it e.g.},][]{1984JAP....55..764S,1998MNRAS.301..955D,2005A&A...435..885P,2009ASPC..414..473J,2012A&A...548A..68P,2013A&A...xxx...xxx}. 

Based on the above and preceding work \citep{1990QJRAS..31..567J,1990MNRAS.247..305J,2009ASPC..414..473J,2012aA&A...540A...1J,2012bA&A...540A...2J,2012cA&A...542A..98J,2013A&A...xxx...xxx,2013A&A...555A..39J} the principal processes in the a-C(:H) dust evolutionary cycle in the ISM can be summarised as follows: 
\begin{itemize}
\item Dust formation in AGB dust shells  leads to the formation of wide band gap ($E_{\rm g} \sim 2.5$\,eV), aliphatic-rich a-C:H dust, which is aromatised in the transition to the ISM (small grains and the outer surfaces of large grains are converted to low band gap a-C, $E_{\rm g} \sim 0$\,eV). 
\item Dust eroded in the ISM (through the effects of EUV photo-processing, shocks and cosmic rays) is then re-accreted in the denser, molecular ISM (A$_{\rm V} > 1$) to form wide band gap ($E_{\rm g} \gtrsim 2$\,eV) mantles on all grains.  Such accreted mantles may be the glue that cements the grains in aggregates and  could also be the source of the observed cloud/core shine \citep[see][and references therein]{2013A&A...xxx...xxx}. 
\item During a-C:H dust (re-)formation by accretion in denser regions hetero-atoms could be incorporated into the structure to form doped materials, {\it i.e.}, a-C:H:$X$, with the element $X$ showing a preference for the aromatic phase and/or aliphatic phase depending upon its chemical proclivity. 
\item Carbon dust evolution, via EUV-UV ($h\nu \geqslant 10$\,eV) photo-processing, in the ISM/H{\footnotesize II} regions/PDRs leads to associated de-hydrogenation, aromatisation and a decrease in band gap (all manifestations of the chemical evolution of the  structure), and can also lead to photo-fragmentation. 
\item EUV-UV photo-fragmentation in intense radiation fields, perhaps aided by grain-grain collisional fragmentation  in turbulent regions of the ISM, is a trickle-down process that leads to a wide range in the a-C(:H) grain size  distribution, which includes:  
\begin{itemize}
\item FUV extinction carriers \ \        ($a \sim 3$\,nm, \ \, a few    1000's of C atoms),  
\item UV bump carriers \ \ \ \ \ \ \ \ \, ($a \sim 1$\,nm, \ \, a few \   100's of C atoms),  
\item emission band carriers \ \ \ \    ($a \sim 0.5$\,nm,   a few \ \, 10's of C atoms),   
\item molecules/radicals  (C$_2$, CN, CCH, l-C$_3$, l-C$_3$H$^+$, c-C$_3$H$_2$, l-C$_4$H, \ldots\ l/c-C$_n$H$_m$,  $m \lesssim n$). 
\end{itemize}
\item Photo-fragmentation of a-C(:H) grains yields aromatic-rich fragments or sub-grains with 10-1000 C atoms, and molecules and radicals with $< 10$ C atoms. The resulting fragment distribution will be incorporated into the ambient ISM where, in comparison, it will be subject to only relatively minor modifications.  
\end{itemize}
The size distribution determined by the photo-physics in H{\footnotesize II} regions and PDRs then determines and sets the small carbonaceous grain size distribution in the diffuse ISM, where the radiation field and hence the effects of photo-processing are significantly reduced. 

Among the smallest a-C(:H) species, containing less than a hundred carbon atoms, there will be a rather wide isomeric variation, which is nevertheless limited by the bonding configuration  in arophatic clusters and is therefore probably based upon a limited set of common `backbone' species.

\section{The Red Rectangle: A case in particular} 
\label{sect_RR}

The Red Rectangle (RR) is an interesting object because it allows us to qualitatively test some of the EUV-UV photo-processing aspects outlined above. 
The RR is a proto-planetary nebula that exhibits a spectacular display of red luminescence, the extended red emission (ERE, $E \simeq 1.8-2.5$\,eV, $\lambda \simeq 500-700$\,nm, peaking at $E \simeq 1.94$\,eV, $\lambda \simeq 640$\,nm) close to the central star (HD\,44179) and a more extended blue luminescence (BL, $E \simeq 3.1-3.5$\,eV, $\lambda \simeq 350-400$\,nm) showing a strong peak at $\sim 3.3$\,eV (375\,nm) and a broad shoulder at $\sim 3.2$\,eV (390\,nm) \citep{2004ApJ...606L..65V,2005ApJ...633..262V,2005IAUS..235P.234V}. 
Narrowband images of the RR show that the morphology of the BL and ERE are essentially mutually exclusive, with the BL coincident with the outermost parts of the circumstellar disc and the ERE localised in and within the walls of the outflow cavities \citep{2006ApJ...653.1336V}.
Further, \cite{1989ASS...150..387D} showed that some of the sharp emission lines superimposed on the ERE bear a striking resemblance to the zero phonon lines of terrestrial diamonds, indicating that $sp^3$-rich, wide band gap a-C:H dust is an important dust component in the RR nebula.
In the RR the ERE does not appear to be strongly associated with the IR emission bands but it does correlate with FUV photons ($E > 10.5$\,eV), 
in that the ERE needs $> 10$\,eV photons for excitation \citep{1985ApJ...294..225W,2006ApJ...636..303W}. 
This is the same EUV-UV photon energy that is required to de-hydrogenate and narrow the band gap in a-C:H materials \cite{2012aA&A...540A...1J,2012bA&A...540A...2J,2012cA&A...542A..98J}. 
The laboratory experiments and analysis of \cite{2010A&A...519A..39G} show that the ERE is indeed consistent with a-C:H materials. However, the photo-luminescent (PL) band width and the PL efficiency measured by  \cite{2010A&A...519A..39G} do not agree with ERE observations, which indicate a narrower range of PL band widths and PL efficiencies that are much higher than those measured in the laboratory for a-C:H thin films ({\it i.e.}, bulk materials). Given that particle size is such a key parameter in determining the optical and physical properties of a-C:H \citep{2012cA&A...542A..98J} it is likely that the PL characteristics will be size-dependent and that therein may lie the source of the discrepancy between the observations and the laboratory data.

As shown by \cite{2009ApJ...693.1946W} the RR system has an interior Lyman/far-UV continuum originating from an accretion disk surrounding the main sequence secondary star, which is fed by Roche lobe overflow from the post-AGB primary. 
The central system is surrounded by an optically thick, edge-on, circumstellar disc, which extinguishes forward-scattered radiation \citep[{\it e.g.},][]{2006ApJ...653.1336V}.
Thus, the geometry of the RR region is rather unusual, in that there are both internal and external sources of UV photons, the external UV photons coming from the ambient interstellar radiation field.  
The ERE only occurs inside the outflow cavity and along its walls where there is an unobscured view of the FUV photons emanating from the central stars \citep{2006ApJ...653.1336V}. 
Whereas, the BL is only associated with those regions that are most shielded from the interior FUV radiation, {\it i.e.}, the outermost parts of the disc.
With this configuration it is possible to explain the interior ERE and extended BL spatial structure in the RR.
In the inner regions the UV continuum photons will transform any wide band gap a-C:H ($E_{\rm g} \gtrsim 2$\,eV) into narrower band gap a-C ($E_{\rm g} < 1$\,eV). The innermost carbonaceous dust is thus re-configured into an ERE carrier by the UV photons, which also excite it. Further out the wide band gap a-C:H ($E_{\rm g} \gtrsim 2$\,eV) dust in the 
outer regions of the circumstellar disc 
is not transformed and it is this wider band gap, aliphatic-rich material that is responsible for the higher energy BL. 
The BL being more extended and shielded from the Lyman/far-UV continuum from the centre of the RR system implies that the a-C:H carriers are not  photo-processed into a-C ERE carriers. It is also possible that interstellar UV photons could be exciting the BL carriers in the outer disc regions.
Eventually, and after injection into the ambient ISM, the a-C:H to a-C transformation will occur over timescales of the order of $10^5-10^6$\,yr  \citep{2012cA&A...542A..98J}. Hence, the BL does not extend far beyond the bounds of the RR region because further out it is, or has been, transformed into an ERE carrier or perhaps the interstellar radiation field is too weak to excite it. 


The BL does correlate with the 3.3\,$\mu$m emission band, which is generally assumed to be due to small carbonaceous particles with a few 10's of C atoms \citep{2005ApJ...633..262V}. 
However, it is clear that the BL may not only be due to very small particles because, 
as pointed out by \cite{2013A&A...555A..39J},
a similar luminescence is also observed in taC:H and ta-C:N thin films in the laboratory ($E_{\rm g} = 1-2.5$\,eV), which  show a broad blue luminescence ($E \sim 2.2-3.7$\,eV, $\lambda \sim 335-654$\,nm) with a peak at 2.66\,eV and other superimposed bands \citep{2006TSF...515.1597P}.  
It is interesting that one of the weaker superimposed bands in the taC:N luminescence occurs at 442.8\,nm, the same wavelength as the strongest DIB. 
Thus, given the RR observations, the photo-processing of a-C:H materials could provide viable DIB carrier candidates. 
Indeed, a photo-processing scheme has been suggested for the formation of the 580\,nm emission line carrier in the RR, which may be related to the 579.7\,nm DIB \citep{1998MNRAS.301..955D}. 
It has also been noted that the carrier of the 579.7\,nm DIB towards $\chi$\,Vel  is more sensitive to the interstellar UV radiation field than to the local density \citep{2013MNRAS.429..939S}.
Nevertheless, and despite the strong evidence for a-C:H photo-processing in the RR, only two weak DIBs (578.0 and 661.3\,nm) have been observed in absorption in the spectrum along this line of sight, originating in either in the RR region itself or in an intervening diffuse interstellar cloud \citep{2004ApJ...615..947H}. Thus, and because of the high local extinction ($A_{\rm V} \sim 4$\,mag.), it seems that the RR probably forms the DIB precursors and that the actual DIB carriers are only produced by further UV photo-processing in the low-density diffuse ISM.

\section{The DIB connection}
\label{sect_connection}

\begin{figure}
\begin{center}
\includegraphics[width=10.0cm,angle=90]{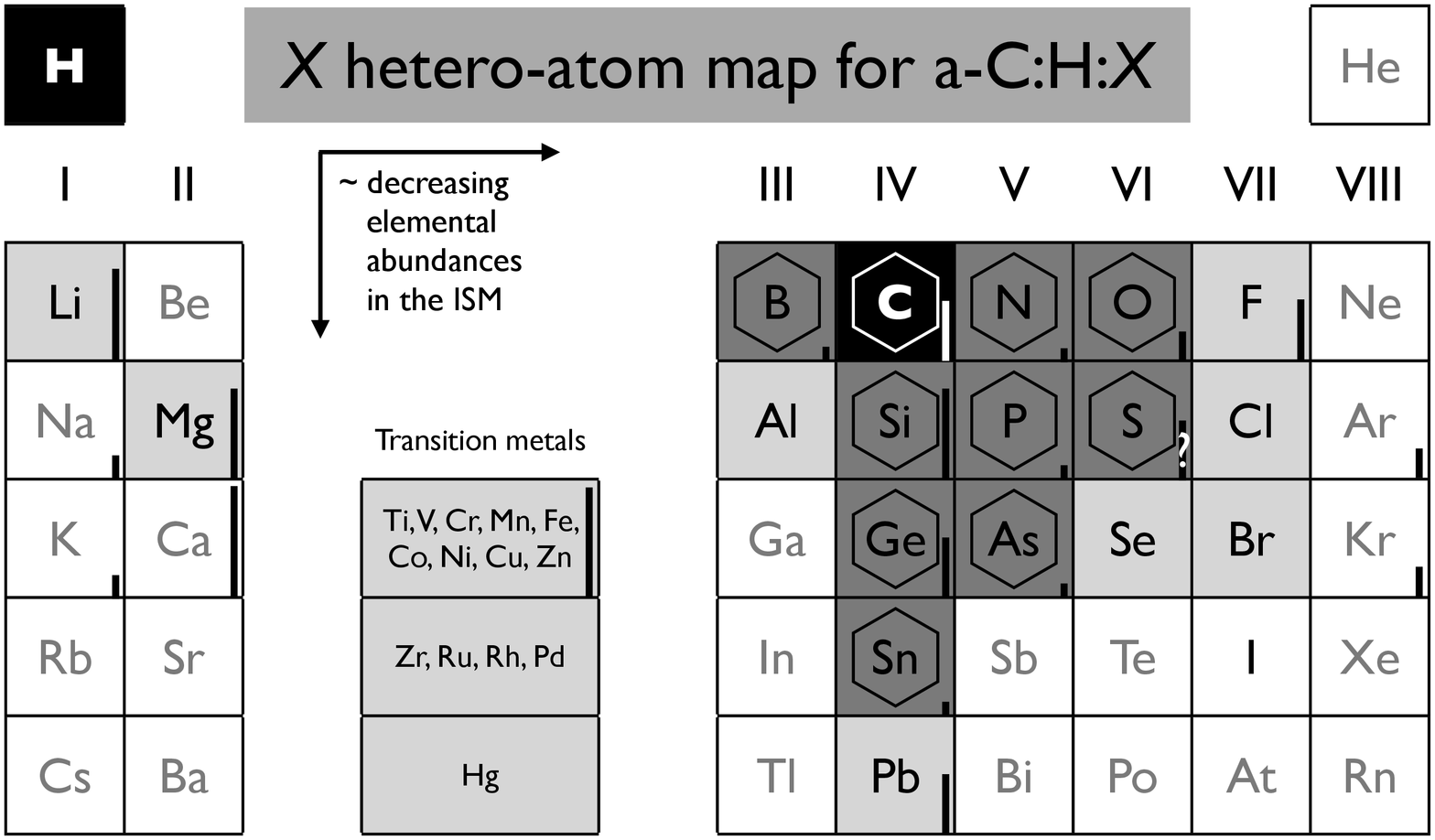}
\caption{A hetero-atom periodic table map for  a-C:H:{\it X} materials.  The elements shaded dark-grey and encircled by hexagons can incorporate into 5- and 6-membered aliphatic and aromatic hetero-cyclic compounds. The elements in the light-shaded boxes form known organic and organometallic compounds, and include many of the transition metals. The bar on the right hand side of a box gives an indication, where known, of the elemental depletion in the ISM (where a bar from bottom to top would imply 100\% depletion of the element into dust). Note that the ISM structure-tracing elements Na, K and Ca do not appear to be DIB-related}
\label{fig_DIB_PT}
\end{center}
\end{figure}

As pointed out in \cite{2013A&A...555A..39J} there may be a rather interesting connection between the hetero-atom doping of a-C:H materials and several depletion conundrums, luminescence from dust and the DIBs. Indeed, based upon the many studies of the Red Rectangle region, there appears to be an intriguing link between the ERE, BL and the DIB-like emission bands seen in this object. Of particular interest is the fact that this object shows BL and that in the laboratory a blue luminescence in ta-C:N materials shows a 442.8\,nm band at the same wavelength as the most prominent DIB. 

Thus, it appears that hetero-atom doping of a-C:H, be it N as above or other hetero-atoms (most likely at the percent level), could produce well-defined optical bands in amorphous hydrocarbon materials. 
Indeed, single hetero-atoms, {\it e.g.}, of B, N, O, Si, P, S, Ge, As and Sn, can be incorporated into organic molecules in 5- and 6-membered aliphatic and aromatic cyclic compounds.
Additionally, other elements such as Li, Mg, Se, Pb and many transition metals are important in organometallic chemistry.  
All of this leads to a very rich chemistry for a-C:H hetero-atom doping and the possibility of a whole host of interesting optical bands. 
With this in mind Fig.~\ref{fig_DIB_PT} gives a summary of the hetero-atom chemistry of carbon, in terms of both the hetero-cycle and organometallic chemistry.  
In this figure it is noteworthy that the elements Na, K and Ca,  
which are often used to trace ISM structure, do not appear to be incorporable into a-C:H and are therefore predicted to be unrelated to the DIBs within the framework proposed here. 
However, DIBs do show a  power law dependence on Na{\footnotesize I}, K{\footnotesize I} and C{\footnotesize I} \citep{1995ARA&A..33...19H}. 
The halogens, and in particular Cl and F could also be active a-C(:H) dopants because of their very active chemistry. 

Most of the elements, outside of C, N, O, Si, P, S (the 1$^{\rm st}$ and 2$^{\rm nd}$ row elements in groups IV, V and VI of the periodic table), are only trace elements in the ISM and can therefore only be present as trace hetero-atoms in  a-C:H:$X$ but this may be enough. 
However, given the cosmic abundances and the observed elemental depletions, the most likely a-C(:H) dopants, 
at the $\gtrsim$1\% level, would appear to be O, N, Mg, Si, Fe, S and Ni, while P, Cl, Cr and Mn could only be present at the sub-1\% level even if they were to be completely depleted into a-C:H, an unlikely scenario. Other possible dopants could, at most, only be present at the sub 0.1\% level. 
Hence, these low cosmic abundance elements could only contribute to the DIBs if the oscillator strengths for the relevant transitions were extremely strong.  
This perhaps explain why Ti does not correlate with the DIBs \citep{1995ARA&A..33...19H} and indicates that low abundance transition metals (other than Fe and Ni, and perhaps Cr and Mn) might not be associated with the DIB carriers. 

It should be noted that to date \citep[with the exception 619.60 and 661.36\,nm DIBs,][]{2010ApJ...708.1628M} no clear correlations between any DIBs has yet been found and there also appears to be no unique DIB spectrum \citep[{\it e.g.},][]{2009ApJ...705...32H,2011ApJ...727...33F}, all of which strongly suggests that each DIB is essentially unique in origin.

As can perhaps be surmised from Fig.~\ref{fig_DIB_PT}, even if a hetero-atom only gives rise to one optical band then a number of completely independent bands are possible. However, it is most likely that a given hetero-atom can be incorporated into a number of different bonding sites, each with a characteristic but different optical band in the numerous isomeric forms of the a-C:H sub-nm particles. Hence, and on the basis of the scenario proposed here, each $X$ hetero-atom in a-C:H:$X$ is likely to yield, at least, many tens of independent optical bands, which will be uncorrelated. In addition, the sub-components of the a-C:H structure in un-doped materials, or the un-doped parts of the structure in a-C:H:$X$, may also produce their own characteristic optical bands.

During a-C:H(:$X$) dust (re-)formation, in the denser molecular regions of the ISM, $X$ hetero-atoms of O, N, Mg, Si, Fe, S, Ni,  \ldots\ could be incorporated into the a-C:H structure with some elements showing a chemical preference for the $sp^2$ olefinic/aromatic phase and others a preference for the $sp^3$ aliphatic phase \citep{2013A&A...555A..39J}. It is perhaps to be expected that the hetero-atoms with a preference for the aromatic phase will be more resistant to EUV-UV photo-processing than those incorporated into an aliphatic phase \citep[{\it e.g.},][]{2013A&A...555A..39J}. Given that some elements can incorporate into either phase some overlap in the sensitivity to UV processing is therefore inevitable. 

It is therefore likely that, in the a-C:H:{\it X} interpretation of DIBs, any optical band will depend upon the hetero-atom incorporation site rather than on particle size because the latter would lead to wide variations in the width of a given DIB. 
However, the broader DIBs that are less sensitive to UV radiation could arise from larger and/or more resistant aromatic-rich a-C particles. 
Nevertheless, particular DIBs ought to strongly reflect the particular chemical and structural nature of the site where the hetero-atoms are actually incorporated, as well as their abundance in those sites, as required by the stability of the DIB peak positions. 
Given that the $d$ orbital configurations of the transition metals produce diagnostic spectrophotometric bands in the visible, it would be interesting to investigate their effects on a-C:H spectra. 


\section{Suggestions for a concerted approach to the DIB problem}
\label{sect_attack}

In ISM elemental depletion studies it would perhaps be worthwhile to look for correlations between the DIBs and the most likely hetero-atom dopants of a-C:H, {\it i.e.}, O, N, Mg, Si, Fe, S and Ni, as indicated in Fig.~\ref{fig_DIB_PT}.  It would seem that the best candidate elements for DIB depletion correlation studies ought to be the more abundant elements that show only relatively small depletions and that therefore only weakly correlate with dust extinction, {\it i.e.}, N and S. Elements such as O, Si, Mg, Fe and Ni are already predominantly in dust, which already correlates with the DIBs. However, that fraction of the element that remains in the gas could still show anti-correlations with the DIBs. 

From the above it is clear that it would be of fundamental interest to undertake a detailed laboratory study of the visible-near infrared spectra of doped a-C:H materials, a-C:H:{\it X}, where $X$ may be B, N, O, Si, P, S, Ge, As, Sn, Li, Mg, Se, Pb and transition metals and to look for DIB-like bands. Such an undertaking would be a huge effort but it might reveal the underlying family ties between the numerous DIBs. It would seem to be that focussing on a-C(:H) doped with N, S and Si hetero-atoms, and also on fullerenes doped with these elements, could potentially be the most promising and profitable starting points \cite[{\it e.g.},][]{2013A&A...555A..39J}. 

As a follow-up to the above, and in order to understand their resistance to photo-processing, the EUV-UV irradiation of a-C:H and a-C:H:{\it X} materials and the analysis  of their `free-flying' dissociation/decomposition products, coupled with mass spectrometry and visible/near-infrared spectroscopy could perhaps help to shed some light on the possible nature of the DIB carriers and their evolution in the ISM and in circumstellar regions under the effects of EUV-UV irradiation. 
Indeed, and in addition to $X$-doped a-C:H nano-particles, it is also possible that some of the daughter products of the EUV-UV photo-fragmentation of a-C(:H) nano-particles \citep[{\it i.e.}, small linear/cyclic species  l/c-C$_n$H$_m$, where $n=3,4,$ \ldots\  and $m \lesssim n$,][]{2012aA&A...540A...1J,2012cA&A...542A..98J} could also be potential DIB carriers.  
This is perhaps indicated by the coincidence between the bands of l-C3H2 and several DIBs 
\citep{2011ApJ...726...41M,2012JChPh.136m4312S}, 
even though this specific radical may not be the main contributor to these DIBs 
\citep{2011Sci...331..293O,2011ApJ...735..124K,2012ApJ...753L..28L,2012ApJ...753L..11A}.

\section{Conclusions}
\label{sect_conclusions}

It is concluded from this study that the FUV extinction carriers and the UV bump carriers are not the same as the DIB carriers but that the latter are the photo-processing products of the larger particles responsible for the former. 
Thus, the FUV extinction, UV bump and DIB carriers may be directly related through photo-processing, in that the DIB carriers are the photo-fragmentation products of the a-C:H(:$X$) nano-particles that are responsible for the UV extinction features. The DIBs then ought, as observed, to be associated with the diffuse ISM and be weaker where they themselves are photo-fragmented into molecular, radical and ionic daughter species (C$_2$, CN, CCH, C$_3$, C$_3$H$^+$, c-C$_3$H$_2$, C$_4$H, \ldots) in intense radiation fields, such as in PDRs. Hence, the UV bump and DIBs are expected to be weaker where the UV radiation field is intense and therefore where the photo-dissociation daughter molecules, radicals and ions are abundant. 

It is here proposed that DIBs arise from hetero-atom doping `colour centres' in a-C:H:$X$ ($X=$ O, N, Mg, Si, Fe, S, Ni, P, Cl, Cr and Mn ) sub-nano-particles with fewer than 100 carbon atoms, which likely have a common backbone structure rather similar to arophatic species. As observed these `colour centre' bands ought to be broader than molecular lines and will clearly be environment dependent. 

It is also highly likely that some of the larger, non-hetero-atom doped, photo-fragmentation products of a-C(:H), {\it i.e.}, linear and cyclic species, l/c-C$_n$H$_m$, $n \geqslant 3$ and $m \lesssim n$, could also be viable DIB carriers. 

Thus, while this paper certainly does not provide a solution to the long-standing problem of the nature and origin of the DIB carriers, it does suggest a new view and a rather promising framework within which a resolution to the DIB carrier  problem may perhaps be found.

\section{Acknowledgements}
\label{sect_acknow}

The author would especially like to thank Hiroshi Kimura and Akio Inoue, and indeed all of the organisers of ``Cosmic Dust VI'' held in Kobe in August 2013, for their help and for a most interesting and scientifically-stimulating meeting. 
This research was, in part, made possible through the financial support of the Agence National de la Recherche (ANR) through the programme CIMMES (ANR-11-BS56-029-02).

\bibliographystyle{elsarticle-harv}
\bibliography{biblio_HAC.bib}







\end{document}